\documentclass[prd,showpacs,superscriptaddress,twocolumn,nofootinbib,floatfix,10pt]{revtex4-2}
\usepackage[utf8]{inputenc}
\usepackage{amsmath,amssymb}
\usepackage{slashed}
\usepackage{subfigure}
\usepackage[colorlinks=true,
breaklinks=true,
urlcolor=magenta,
citecolor=blue]{hyperref}
\usepackage[usenames,dvipsnames]{color}
\usepackage{appendix}
\usepackage{braket,bm}
\usepackage{multirow}
\usepackage{enumitem}
\usepackage{color}
\usepackage{orcidlink}
\usepackage{graphicx}
\usepackage{tikz}
\usepackage{booktabs}
\usepackage{array}
\usepackage{overpic}
\usepackage{float}
\usepackage[normalem]{ulem}
\usepackage{threeparttable}
\allowdisplaybreaks[4]

\graphicspath{{figs/}}
\newcommand{\al}{&\!\!\!\!}
\newcommand{\bea}{\begin{eqnarray}} 
\newcommand{\eea}{\end{eqnarray}}

\newcommand{\BB}{{\mathbb B}}
\newcommand{\PP}{\mathcal P\bar{\mathcal P}}
\newcommand{\JN}{J/\psi N}

\newcommand{\itp}{\affiliation{CAS Key Laboratory of Theoretical Physics, Institute of Theoretical Physics,\\ Chinese Academy of Sciences, Beijing 100190, China}}

\newcommand{\ucas}{\affiliation{School of Physical Sciences, University of Chinese Academy of Sciences, Beijing 100049, China}}

\newcommand{\peng}{\affiliation{Peng Huanwu Collaborative Center for Research and Education, Beihang University, Beijing 100191, China}}

\newcommand{\hiskp}{\affiliation{Helmholtz-Institut f\"{u}r Strahlen- und Kernphysik and Bethe Center for
Theoretical Physics, \\Universit\"{a}t~Bonn,  D-53115~Bonn,~Germany}}

\newcommand{\uestc}{\affiliation{School of Physics, University of Electronic Science and Technology of China, Chengdu 611731, China}}

\newcommand{\scnt}{\affiliation{Southern Center for Nuclear-Science Theory (SCNT), Institute of Modern Physics,\\ Chinese Academy of Sciences, Huizhou 516000, China}}

\newcommand{\thu}{\affiliation{Department of Physics, Tsinghua University, Beijing 100084, China}}

\usepackage{graphicx} 
\makeatletter
\def\@fnsymbol#1{\ensuremath{\ifcase#1\or \star\or \dagger\or \ddagger\or
   \mathsection\or \mathparagraph\or \|\or **\or \dagger\dagger\or \ddagger\ddagger\else\@ctrerr\fi}}
\def\fps@figure{tbp}
\def\ftype@figure{1}
\def\ext@figure{lof}
\def\fnum@figure{\textbf{Fig.~\thefigure}} 
\makeatother

\begin{document}
\title{Deciphering the mechanism of $J/\psi$-nucleon scattering} 
\author{Bing Wu\orcidlink{0009-0004-8178-3015}}
\uestc\itp\ucas
\author{Xiang-Kun Dong\orcidlink{0000-0001-6392-7143}}
\hiskp
\author{Meng-Lin Du\orcidlink{0000-0002-7504-3107}}\email{du.ml@uestc.edu.cn (M.L. Du)}
\uestc

\author{Feng-Kun Guo\orcidlink{0000-0002-2919-2064}}\email{fkguo@itp.ac.cn (F.K. Guo)}
\itp\ucas\peng \scnt

\author{Bing-Song~Zou\orcidlink{0000-0002-3000-7540}}\thu 

\begin{abstract}

\textbf{{Abstract:}} The low-energy $J/\psi N$ scattering is important for various reasons: it is related to the hidden-charm $P_c$ pentaquark states, provides insights into the role of gluons in nucleon structures, and is relevant to the $J/\psi$ properties in nuclear medium.
The scattering can happen through two distinct mechanisms: the coupled-channel mechanism via open-charm meson-baryon intermediate states, and the soft-gluon exchange mechanism.
We investigate the $J/\psi N$ $S$-wave scattering length through both mechanisms, and find that the soft-gluon exchange mechanism leads to a scattering length at least one order of magnitude larger than that from the coupled-channel mechanism and thus is the predominant one. 
The findings can be verified by lattice calculations and will enhance our understanding of the scattering processes breaking the Okubo-Zweig-Iizuka rule. 

\vspace{10pt}
\textbf{\textit{Keywords:}} Hadron-hadron scattering; Dispersion relation; Nucleon structure; Charmonium; Okubo-Zweig-Iizuka (OZI) rule

\end{abstract}

\maketitle

\section{Introduction}

There has been sustained interest in the low-energy $\JN$ interaction for a long time~\cite{Peskin:1979va,Bhanot:1979vb}, as it can offer vital insights into the role of gluons in the hadron structures and interactions among hadrons, which are still challenging problems of quantum chromodynamics (QCD) in the nonperturbative region. It is crucial in describing the photo- and hadro-production of charmonia on the nuclear targets, which is of increasing interest, especially after the discovery of pentaquark $P_c$ states by the LHCb Collaboration~\cite{LHCb:2015yax,LHCb:2019kea} and due to the possibility of probing the trace anomaly contribution to the nucleon mass~\cite{Kharzeev:1995ij, Kharzeev:1998bz}.
It is also relevant to the properties of the $J/\psi$ in nuclear medium and possible $J/\psi$-nucleus bound states~\cite{Luke:1992tm, Sibirtsev:2005ex, Wu:2012wta} and to identifying signals of the quark-gluon plasma~\cite{Barnes:2003vt}.
Such processes will be explored in future experiments at the Electron-Ion Collider (EIC)~\cite{Burkert:2022hjz}, the possible 22~GeV upgrade at the Jefferson Laboratory~\cite{Accardi:2023chb} and the Electron-Ion Collider in China~\cite{Anderle:2021wcy}.

Since $J/\psi$ and nucleon have no valence quarks in common, their scattering violates the Okubo-Zweig-Iizuka (OZI) rule~\cite{Okubo:1963fa,Zweig:1964jf,Iizuka:1966fk} and the amplitude is suppressed by $1/N_c$ compared to the ones for OZI allowed meson-baryon scattering processes, with $N_c$ being the number of colors in QCD~\cite{tHooft:1973alw}. 
The $\JN$ scattering, and in general OZI suppressed scattering processes, can proceed through two distinct mechanisms. The first involves the exchange of multiple gluons, which is usually treated using the QCD multipole expansion method and results in the so-called gluonic van der Waals interactions~\cite{Peskin:1979va,Bhanot:1979vb,Luke:1992tm,Brodsky:1989jd,Kaidalov:1992hd,deTeramond:1997ny,Brodsky:1997gh,Sibirtsev:2005ex,Beane:2014sda,Krein:2020yor}. 
It has two building blocks, namely, the charmonium-gluon coupling in terms of the chromopolarizability that describes the ability of the charmonium to emit gluons~\cite{Voloshin:1979uv,Leutwyler:1980tn,Fujii:1999xn,Guo:2006ya,Brambilla:2015rqa,Chen:2019hmz} and the matrix elements of the gluonic operator between the single nucleon states, associated with the trace anomaly contribution to the nucleon mass~\cite{Chanowitz:1972vd,Crewther:1972kn,Voloshin:1980zf}. 
The second mechanism encompasses coupled channels due to the rescattering from the $\JN$ to channels with open-charm meson and baryon states ($\Lambda_c \bar D^{(*)}$ and $\Sigma_c^{(*)}\bar D^{(*)}$)~\cite{Du:2020bqj} and back to $\JN$.\footnote{It has been noticed that meson loops may evade the OZI suppression in the light meson sector with certain $J^{PC}$ quantum numbers~\cite{Lipkin:1996ny}.} 
Therefore, a study of the trace anomaly contribution to the nucleon mass with the $J/\psi$ as a probe is possible only if the first mechanism is dominant, and it is vital to decipher the mechanism of the $J/\psi N$ scattering at low energies. 
This work is devoted to solving this problem.

Despite extensive exploration of the $\JN$ interaction using various methods, the strength of this interaction remains a topic of ongoing debate. Different theoretical studies report significant variations in the $S$-wave $\JN$ scattering length, spanning several orders of magnitude~\cite{deTeramond:1997ny,Brodsky:1997gh,Sakinah:2024cza,Krein:2020yor,Sibirtsev:2005ex,Hayashigaki:1998ey,Shevchenko:1996ch,Du:2020bqj,Pentchev:2020kao}.
In particular, lattice QCD studies have not yielded consistent results~\cite{Yokokawa:2006td,Liu:2008rza,Kawanai:2010ev, Beane:2014sda, Alberti:2016dru,Sugiura:2017vks,Skerbis:2018lew}, with predictions ranging from a strong attraction, leading to a deep $\JN$ bound state obtained with a pion mass about 805~MeV~\cite{Beane:2014sda}, to a very weak $\JN$ interaction~\cite{Skerbis:2018lew}.

One ongoing effort is to extract the $\JN$ scattering length from the near-threshold photoproduction of $J/\psi$ on the proton~\cite{Strakovsky:2019bev,Pentchev:2020kao, GlueX:2023pev}, which is being studied by experiments at the Jefferson Laboratory~\cite{GlueX:2019mkq, GlueX:2023pev, Duran:2022xag}. 
The method is based on the vector-meson dominance model such that the nearly on-shell photon is converted into a $J/\psi$ and then interacts with the proton~\cite{Kharzeev:1998bz}.
However, the $\JN$ scattering length is defined for the on-shell $J/\psi$ and nucleon, while the intermediate $J/\psi$ connected to the photon is highly off-shell.
Furthermore, the applicability of the vector-meson dominance model in the charmonium sector for this process has also been challenged in Refs.~\cite{Du:2020bqj, Xu:2021mju, JointPhysicsAnalysisCenter:2023qgg}.\footnote{ In Ref.~\cite{Sakinah:2024cza}, a phenomenological charm quark-nucleon potential was developed based on a specific constituent quark model~\cite{Segovia:2013wma}, avoiding the use of the vector-meson dominance model, and the Pomeron-exchange model~\cite{Donnachie:1984xq, Wu:2012wta,Wu:2013xma,Lee:2022ymp}. The potential was employed to study the $J/\psi N$ interaction and the photoproduction of $J/\psi$, $\eta_c$ and $\psi(2S)$ on the nucleon. However, the open-charm meson-baryon coupled-channel effects are not included in that study.} In particular, it has been shown in Refs.~\cite{Du:2020bqj,JointPhysicsAnalysisCenter:2023qgg} that the $J/\psi$ photoproduction data can be well described by the coupled-channel mechanism, and the predicted existence of $\Lambda_c\bar D^{(*)}$ threshold cusps~\cite{Du:2020bqj} is in line with the GlueX data~\cite{GlueX:2023pev}.

In this work, we aim at deciphering the dominant mechanism for the $\JN$ scattering by calculating the $S$-wave scattering length in both the coupled-channel and multigluon exchange mechanisms. 
The former is based on the chiral effective field theory framework constructed in Ref.~\cite{Du:2021fmf}, while the latter is evaluated using the technique of dispersion relations.
We will show that at low energies, the $\JN$ interaction is dominated by the soft-gluon exchange mechanism, which results in an attraction that is not yet strong enough to support a near-threshold pole.

\section{Coupled-channel mechanism}

In the coupled-channel mechanism, the $\JN$ interaction is provided by open-charm meson-baryon intermediate states.
The scattering amplitudes for the coupled-channel system, including $\JN$, $\eta_c N$, $\Lambda_c\bar D^{(*)}$ and $\Sigma_c^{(*)}\bar D^{(*)}$, have been constructed by solving the Lippmann-Schwinger equation in the chiral effective field theory framework in Ref.~\cite{Du:2021fmf} (for a recent coupled-channel study using the light meson exchange model, see Ref.~\cite{Zhang:2024dkm}).
The long-range interaction is mediated by the one-pion exchange, and the short-range coupled-channel interactions are parameterized in terms of contact terms. Next-to-leading-order contact terms for the $S$-wave-to-$D$-wave transitions are introduced to render the results cutoff independent.
Heavy quark spin symmetry is employed to relate various couplings involving charmed hadrons in the same spin multiplet.
The parameters are fixed by fitting to the LHCb data on the $J/\psi p$ invariant mass distribution of the $\Lambda_b^0 \rightarrow J / \psi p K^{-}$ decay~\cite{LHCb:2019kea}.  
Details of the framework can be found in Ref.~\cite{Du:2021fmf}.
It is worth noting that the  direct $J/\psi N\to J/\psi N$ scattering amplitude is not included in the coupled-channel framework. Thus, the result can be regarded as due to the open-charm meson-baryon coupled channels only, without the gluon-exchange mechanism to be considered in the next section.

Using the scattering amplitudes obtained therein, we calculate the $S$-wave $\JN$ scattering length, defined via the effective range expansion as:
\begin{equation}
    p \cot\delta_{0,J/\psi N} = -\frac{1}{a_{0,J/\psi N}} + \frac{1}{2} r_{0,J/\psi N}\, p^2 + \cdots,
\end{equation}
where $p$ is the magnitude of the $J/\psi$ momentum in the $\JN$ center-of-mass (c.m.) frame, $\delta_{0,J/\psi N}$, $a_{0,J/\psi N}$ and $r_{0,J/\psi N}$ are the $S$-wave phase shift, scattering length and effective range, respectively.
The result is: 
\begin{align}
    a_{0,J/\psi N}^\text{CC} \in [-10, -0.1]\times 10^{-3}~\text{fm}, 
    \label{eq:acc}
\end{align}
where the large uncertainty comes from that of the parameters from fitting to the LHCb data.
We have used ``CC'' to denote the coupled-channel mechanism.
The result means that the charmed meson-baryon coupled channels lead to a very weak attraction between $J/\psi$ and nucleon.
 
\section{Gluon exchange mechanism}

\begin{figure}[tb]
\centering
    \includegraphics[width=1.0\linewidth]{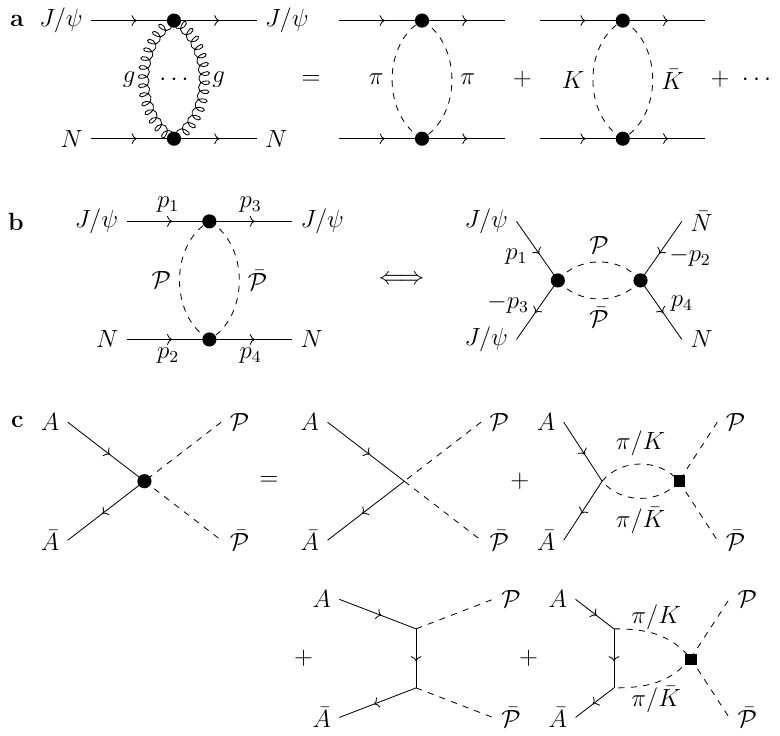}
\caption{\textbf{Feynman diagrams illustrating key steps in the calculation.} (a) Soft-gluon exchange between $J/\psi$ and nucleon, which is equivalent to the exchange of $\pi\pi$, $K\bar K$ and other heavier hadrons. (b) Crossing symmetry between $\JN$ scattering and $J/\psi J/\psi\to N\bar N$. (c) $A\bar A\to\PP$ ($A=N,J/\psi$, and $\mathcal P=\pi,K$) amplitude with the $\pi\pi$-$K\bar{K}$ coupled-channel rescattering, where the black dot represents the full $A\bar A\to\PP$ amplitude, and the black square denotes the $\pi\pi$-$K\bar{K}$ coupled-channel rescattering.}
  \label{fig:feyndiag}
\end{figure}

Because the exchanged gluons between $J/\psi$ and nucleon must be in a color-singlet and isospin-singlet combination, their effects can be captured by inserting a complete set of hadrons with the same quantum numbers as the exchanged multiple gluons.
The most important contribution comes from the lightest $\pi\pi$ exchange, as well as the $K\bar K$ exchange due to the strong $\pi\pi$-$K\bar K$ coupling, as depicted in Fig.~\ref{fig:feyndiag}a.\footnote{The $\eta\eta$ channel is known to be insignificant in the light scalar sector~\cite{Kaiser:1998fi}.}

The exchange of soft gluons with the vacuum quantum numbers can be evaluated using the technique of dispersion relations, which has been employed to show that the exchange of correlated $S$-wave pion pair can be parameterized in terms of a scalar sigma exchange~\cite{Donoghue:2006rg,Wu:2023uva}. 
The $\JN$ scattering amplitude from the $\pi\pi$ and $K\bar K$ (denoted by $\PP$ in the following with $\mathcal P=\pi,K$) exchanges is related to the one of $J/\psi J/\psi\to \PP \to N\bar N$ by crossing symmetry as shown in Fig.~\ref{fig:feyndiag}b. The latter can be calculated through a dispersion relation as:
\begin{align}
    &\mathcal{T}_{0,J/\psi J/\psi \to N(\lambda_3)\bar{N}(\lambda_4)}(s)= \sum_{\mathcal P=\pi,K} \frac{\lambda_3+\lambda_4}{\pi} \notag\\
    & \times \int_{4M_\pi^2}^{+\infty}{\rm d}s'\frac{{T}_{0,J/\psi J/\psi\to\PP}(s')\,
\rho_{\mathcal P}^{}(s')\, \mathcal{T}_{0,N\bar{N}\to\PP}^{*}(s')}{ s'-s-i\epsilon}, \label{TJpsiJpsitoNbarNinmaintext}
\end{align}
where $s$ represents the square of the total energy of $\PP$ in their c.m. frame, $\lambda_3$ ($\lambda_4$) denotes the third component of the helicity of $N$ ($\bar N$), $\rho_{\mathcal{P}}(s)=\theta\left( \sqrt{s}-2M_\mathcal{P}\right)\sqrt{1-4M_\mathcal{P}^2/s}/(16\pi)$
is the phase space factor of $\PP$ with $\theta(s)$ the Heaviside step function, and $T_{0,A\bar A\to \PP}(s)$ is the scattering amplitude for the process $A\bar A\to \PP$ with $A=N$ or $J/\psi$. Here $\mathcal{T}_{0,N \bar{N} \to \PP}(s)=T_{0,N \bar{N} \to \PP}(s)/\sqrt{s-4m_N^2}$ is introduced to avoid the kinematic singularity due to partial-wave projection~\cite{Martin:1970hmp,Wu:2023uva}. 
The subscript $0$ indicates that $\PP$ is in an $S$ wave; since this is always the case, this subscript will be omitted in the following for simplicity.

\begin{figure}[tb]
\centering
    \includegraphics[width=\linewidth]{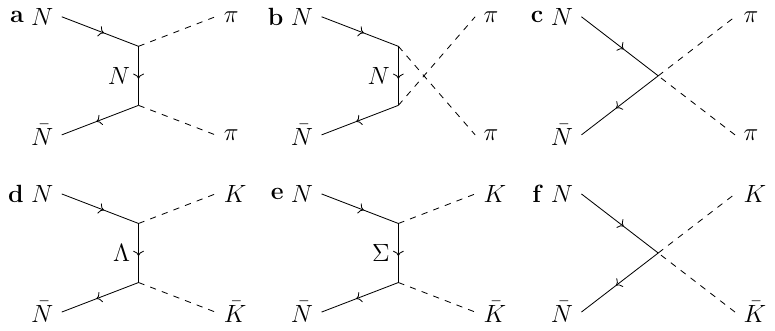}
\caption{\textbf{Tree-level Feynman diagrams for the \boldmath$N\bar{N}\to\pi\pi$ and \boldmath$N\bar{N}\to K\bar{K}$ processes.}}
  \label{FIG2}
\end{figure}

We include the interaction in the $\pi\pi$-$K\bar K$ coupled-channel system shown in Fig.~\ref{fig:feyndiag}c, which is significant and contains the influences of scalar $\sigma$ and $f_0(980)$ exchanges, employing the Muskhelishvili-Omn\`es representation~\cite{Babelon:1976ww,Morgan:1991zx,Garcia-Martin:2010kyn,Hoferichter:2012wf,Lin:2022dyu}.
The procedure has been outlined in Ref.~\cite{Dong:2021lkh}, where the interaction between a pair of $J/\psi$ through exchanging soft gluons was shown to be attractive.
We have:
\begin{align}
\vec{\mathcal{M}}_{A\bar{A}}(s)&=\vec{L}_{A\bar{A}}(s)+\Omega(s)\Biggl[\vec{P}^{(n-1)}_{A\bar{A}}(s)\notag\\
&\quad-\frac{s^n}{\pi}\int_{4M_\pi^2}^{+\infty}{\rm{d}}z\frac{{\rm Im}\left[ \Omega^{-1}(z)\right]}{(z-s)z^n}\Theta(z)\vec{L}_{A\bar{A}}(z)
    \Biggr],\label{equMOrepresentation}
\end{align}
with
\begin{align*}
\vec{\mathcal{M}}_{A\bar{A}}(s)&=\left(\mathcal{M}_{A\bar{A}\to\pi\pi}(s),
    \mathcal{M}_{A\bar{A}\to K\bar{K}}(s)
   \right)^{\rm T},\\
\vec{L}_{A\bar{A}}(s)&=\left(L_{A\bar{A}\to\pi\pi}(s),
    L_{A\bar{A}\to K\bar{K}}(s)\right)^{\rm T},\\
    \vec{P}^{(n-1)}_{A\bar{A}}(s)&=\left(P^{(n-1)}_{A\bar{A}\to\pi\pi}(s),
     P^{(n-1)}_{A\bar{A}\to K\bar{K}}(s)
   \right)^{\rm T},\\
   \Theta(s)&={\rm diag}\left(\theta\left(s-4M_\pi^2\right), \theta\left(s-4M_K^2\right)
   \right),
\end{align*}
where the elements of $\vec P^{(n-1)}(s)$ are polynomials of order $n-1$, and $\vec L_{A\bar{A}}(s)$ represents the possible contributions that do not have the right-hand cuts due to the on-shellness of the intermediate meson pairs (they can have left-hand cuts). 
$\Omega(s)$ is the Omn\`es matrix for the $S$-wave isoscalar $\mathcal P\mathcal{\bar P}$ coupled-channel interaction.
We will employ the Omn\`es matrix elements extracted in Ref.~\cite{Ropertz:2018stk}, which have been calibrated to align with the $\pi\pi$-$K\bar K$ scattering amplitudes reported in Ref.~\cite{Dai:2014zta} at lower energies. Here $\mathcal{M}_{A\bar{A}\to\PP }(s)= T_{J/\psi J/\psi\to \PP}(s)$ for $J/\psi J/\psi\to \PP$ and $\mathcal{M}_{A\bar{A}\to\PP }(s)= \mathcal{T}_{N \bar{N} \to \PP}(s)$ for $N\bar{N}\to \PP$.

The amplitudes $\vec L_{A\bar A}(s)$, which receive left-hand cut contributions, and the polynomials $\vec P^{(n-1)}_{A\bar{A}}(s)$ can be determined by matching to the tree-level amplitudes derived from chiral Lagrangians.
We employ chiral Lagrangians up to $\mathcal{O}(p^2)$, with $p$ a small momentum scale.
That is, we utilize the baryon-meson chiral Lagrangian up to the next-to-leading order (NLO)~\cite{Krause:1990xc,Frink:2004ic,Oller:2006yh}, as well as the leading order (LO) chiral Lagrangian for the emission of light pseudoscalar mesons from charmonia~\cite{Mannel:1995jt,Chen:2015jgl}.

For $A=N$, the left-hand-cut part $L_{N\bar{N}\to\mathcal P\mathcal{\bar P}}(s)$ corresponds to the processes illustrated in Fig.~\ref{FIG2}a, b, d and e, where the involved meson-baryon couplings are from the LO chiral Lagrangian, and the polynomials are matched to the tree-level amplitudes $\mathcal A_{N\bar{N}\to\mathcal P\mathcal{\bar P}}(s)$ of the processes illustrated in Fig.~\ref{FIG2}c and f. 
The involved low-energy constants (LECs) are taken from Fit~II performed in Ref.~\cite{Ren:2012aj}. 
Moreover, given that the explicit forms of $\mathcal A_{N\bar{N}\to\mathcal P\mathcal{\bar P}}(s)$ are linear polynomials in $s$, it is sufficient to adopt the twice subtracted ($n=2$) dispersion relation~\cite{Wu:2023uva}. 

For $A=J/\psi$, there is no left-hand-cut contribution since no resonance near the $J/\psi\pi$ or $J/\psi K$ threshold exists. 
The polynomials are matched to the tree-level chiral amplitudes given by:
\begin{align}
    &\mathcal A_{J/\psi J/\psi \to \PP}(s)=-\frac{2I_{\mathcal P}}{F_\pi^2}\Biggl\{ c_1^{(11)}\left(s-2M_{\mathcal P}^2\right)+2c_m^{(11)}M_{\mathcal P}^2\notag\\
   &\ +\frac{c_2^{(11)}}{12M_{J/\psi}^2}\left[ s^2+2s(M_{\mathcal P}^2+M_{J/\psi}^2)-8M_{\mathcal P}^2M_{J/\psi}^2 \right] \Biggr\},\label{eq:AJpsiJpsi}
\end{align}
with $I_{\pi}=\sqrt{3/2}$ and $I_{K}=\sqrt{{2}}$ the isospin factors. 
The LECs $c_{1,2,m}^{(11)}$ are related to the chromopolarizability of $J/\psi$, $\alpha^{(11)}$, as~\cite{Voloshin:2007dx, Brambilla:2015rqa}: 
\begin{align}
    c^{(ij)}_1=-\left(4+3\kappa\right)f^{(ij)} ,\ 
    c^{(ij)}_2=12 \kappa f^{(ij)},\
    c^{(ij)}_m=-6f^{(ij)}, 
    \label{eq:cij}
\end{align}
where $f^{(ij)}={\pi^2\sqrt{m_{\psi_i} m_{\psi_j}}F_\pi^2}\alpha^{(ij)}/{\beta_0}$, $\beta_0=({11N_c}-{2 N_f})/{3}$ is the first coefficient of the QCD beta function with $N_c$ and $N_f$ the numbers of colors and light-quark flavors, $F_\pi$ is the pion decay constant, and $\kappa$ is an unknown parameter.
The upper index $(ij)$ is introduced to represent $\psi(iS)$ and $\psi(jS)$ states for the involved charmonia (hence $(11)$ appears in Eq.~\eqref{eq:AJpsiJpsi}).

\begin{figure}[tbh]
    \centering
    \includegraphics[width=\linewidth]{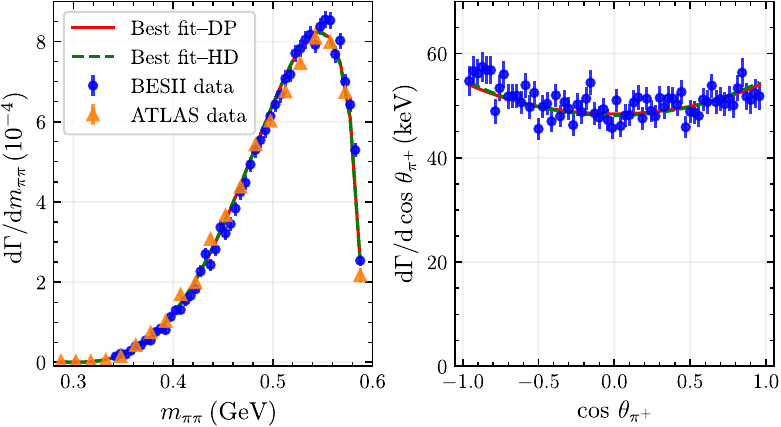}
    \caption{\textbf{Fit to the BESII data~\cite{BES:2006eer} and ATLAS data~\cite{ATLAS:2016kwu} for the \boldmath$\psi(2S)\to J/\psi\pi^+\pi^-$ transition: dipion invariant mass distribution (left) and the helicity angular distribution (right).} The ``Best fit--DP" and ``Best fit--HD" represent the fitting results obtained using the Omn\`es matrix from Ref.~\cite{Ropertz:2018stk} and Ref.~\cite{Hoferichter:2012wf}, respectively.}
    \label{fig:fitall}
\end{figure}
Although the value of the chromopolarizability for the $J/\psi$ determined using various methods~\cite{Eides:2015dtr,Polyakov:2018aey,Brambilla:2015rqa,TarrusCastella:2018php,Ferretti:2018ojb,Dong:2022rwr} bears a quite large uncertainty, the off-diagonal chromopolarizability $\alpha^{(21)}$ can be well fixed using experimental data for the $\psi(2S)\to J/\psi\pi^+\pi^-$ decay~\cite{BES:2006eer,ATLAS:2016kwu}.
Determinations considering the $\pi\pi$ final-state interactions for this transition have been performed in Refs.~\cite{Guo:2006ya,Chen:2019hmz,Dong:2021lkh}.\footnote{The $c_m^{(ij)}$ term was missing in Refs.~\cite{Chen:2019hmz,Dong:2021lkh}.}
Using the Omn\`es matrix from Refs.~\cite{Ropertz:2018stk,Hoferichter:2012wf}, an updated fit to the data of the $\pi\pi$ invariant mass distribution and the helicity angular distribution for the $\psi(2S)\to J/\psi\pi^+\pi^-$ decay~\cite{BES:2006eer,ATLAS:2016kwu} leads to $\kappa=0.26\pm0.01\pm0.01$ and
\begin{align}
    |\alpha^{(21)}|=(1.18\pm0.01\pm0.05) \ \text{GeV}^{-3},
\end{align}
where the first uncertainties come from the errors of the experimental data, while the second ones represent the difference from the Omn\`es matrices in Refs.~\cite{Ropertz:2018stk,Hoferichter:2012wf}.
A comparison of the fit results with the data is shown in Fig.~\ref{fig:fitall}  (see the Appendix for technical details).

Building upon Eq.~(\ref{equMOrepresentation}) and employing the dispersion relation~\eqref{TJpsiJpsitoNbarNinmaintext}, we derive the $s$-channel helicity amplitude $\mathcal{T}_{J/\psi J/\psi \to N(\lambda_3)\bar{N}(\lambda_4)}(s)$ for the process $J/\psi J/\psi\to N\bar{N}$. A Gaussian form factor, $\exp\left[-(s'-s)/\Lambda^2\right]$, is introduced into the integrand of the dispersion integral in Eq.~\eqref{TJpsiJpsitoNbarNinmaintext} to regularize the ultraviolet linear divergence.
This specific form keeps the long-distance behavior of the potential unaffected by the form factor~\cite{Reinert:2017usi}. 
With the crossing relation~\cite{Hara:1970gc,Martin:1970hmp}, this framework enables us to construct the corresponding $t$-channel $S$-wave helicity amplitude for the process of $J/\psi N\to J/\psi N$, as depicted in Fig.~\ref{fig:feyndiag}b, from which one can obtain the $S$-wave $J/\psi N$ scattering amplitude.

\section{Charmonium-nucleon scattering lengths}

We denote the $J/\psi N$ invaratiant mass squared as $t$. At the $J/\psi N$ threshold $t_{\rm th}=(M_{J/\psi}+m_N)^2$, the $S$-wave $J/\psi N$ scattering amplitude reads:
\begin{align}
    T_{0,J/\psi N}(t=t_{\rm th}) &=\al 2m_N \mathcal{T}_{J/\psi J/\psi \to N(\frac{1}{2})\bar{N}(\frac{1}{2})}(s=0),\label{potential2}
\end{align}
which is independent of the total spin $S$, where the dimensional factor $2m_N$ arises from our normalization of the amplitudes in Eq.~\eqref{TJpsiJpsitoNbarNinmaintext}.
The scattering length is given by
$a_{J/\psi N} = -{T_{0,J/\psi N}(t=t_{\rm th})}/{(8\pi\sqrt{t_{\rm th}})}$.

\begin{figure}[tb]
\centering
    \includegraphics[width=0.9\linewidth]{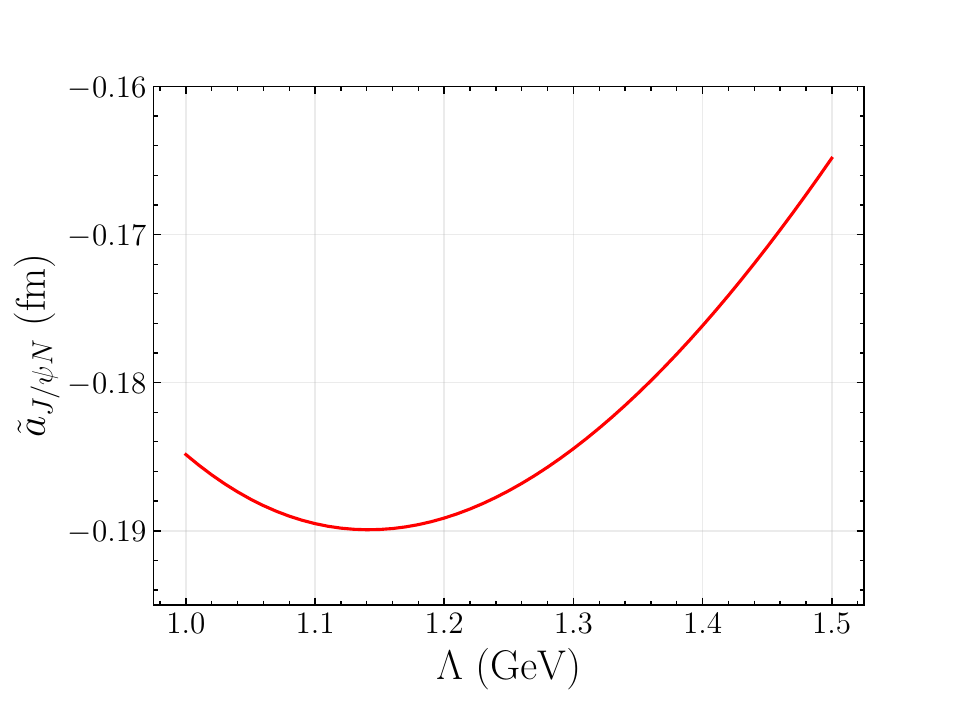}
\caption{\textbf{Dependence of the \boldmath$S$-wave \boldmath$J/\psi N$ scattering length on the cutoff \boldmath$\Lambda$,  evaluated assuming \boldmath$c^{(11)}_{1,2,m}=c^{(21)}_{1,2,m}$. 
}}
  \label{scattringlength}
\end{figure}

To estimate the $J/\psi N$ scattering length, let us first take $c_{1,2,m}^{(11)}=c_{1,2,m}^{(21)}$ to evaluate $\mathcal A_{J/\psi J/\psi \to \PP}$ and the so obtained scattering length will be denoted as $\tilde{a}_{J/\psi N}$.\footnote{The sign of $\alpha^{(11)}$ is chosen to be positive such that it is the same as the estimate for considering the quarkonium as a pure color Coulomb system~\cite{Peskin:1979va}. } 
The result is shown in Fig.~\ref{scattringlength}. 
We find $\tilde a_{J/\psi N}$ falls in the range of $-0.16\sim -0.19$ fm as $\Lambda$ varies from $1.0$~GeV to $1.5$ GeV. The negative sign indicates an attractive interaction. 
Replacing the ${J/\psi}$ mass with the $\psi(2S)$ mass in Eq.~\eqref{eq:AJpsiJpsi},  we obtain $\tilde a_{\psi(2S) N}$ in the range of $-0.14\sim -0.17$~fm.

It has been noticed in Ref.~\cite{Dong:2021lkh} that the absolute value of $\alpha^{(11)}$ should be larger than that of the off-diagonal $\alpha^{(21)}$ because the wave function overlap between two $J/\psi$ states after emitting gluons is larger than that between a $J/\psi$ and a $\psi(2S)$. Thus, we have: 
\begin{equation}
    a_{J/\psi N} \lesssim -0.16~{\rm fm},
\end{equation}
whose absolute value is at least one order of magnitude larger than that from the coupled-channel mechanism given in Eq.~\eqref{eq:acc}.

Indeed, since the LECs $c^{(ij)}_{1,2,m}$ are proportional to $\alpha_{ij}$ as given in Eq.~\eqref{eq:cij}, using the Schwartz inequality~\cite{Sibirtsev:2005ex},
\begin{align}
    \alpha^{(11)}\alpha^{(22)}\geq |\alpha^{(21)}|^2,\label{eq:chromopolarizability}
\end{align}
the product of $\tilde a_{J/\psi N}$ and $\tilde a_{\psi(2S) N}$ derived above in fact sets a lower bound for the product of the $J/\psi N$ and $\psi(2S) N$ scattering lengths, 
\begin{align}
    a_{J/\psi N}a_{\psi(2S) N}\geq \tilde a_{J/\psi N}\tilde a_{\psi(2S) N}\approx {(-0.15\ \rm{ fm}})^2.
\end{align}

\section{Summary}

To summarize, the $J/\psi$ near-threshold photoproduction and the $J/\psi$-nucleon scattering have attracted intensive interest because of their potential connections with measuring the trace anomaly contribution to the nucleon mass.
Thus, it is vital to understand the mechanism behind these reactions. 
There are two possible mechanisms for the low-energy $J/\psi N$ scattering: the mechanism due to charmed meson-baryon coupled channels, and the mechanism due to soft gluon exchanges. 
Only if the latter dominates, it becomes possible to extract the trace anomaly contribution to the nucleon mass from the $J/\psi N$ scattering.
Here we have evaluated the contributions from both mechanisms to the  $S$-wave $J/\psi N$ scattering length.
The contribution from the coupled-channel mechanism is based on the effective field theory formalism in Ref.~\cite{Du:2021fmf}, which has been fixed by reproducing the LHCb data for the $P_c$ hidden-charm pentaquarks.
The contribution from the soft-gluon exchanges was evaluated here by observing that it corresponds to the exchange of light mesons with the same quantum numbers, which was calculated using dispersion relations. We found that both the coupled-channel and soft-gluon exchange mechanisms lead to an attractive interaction, and $[-10,-0.1]\times10^{-3}$~fm and $\lesssim -0.16$~fm, respectively, for the $J/\psi N$ scattering length.
Therefore, we conclude that the low-energy $J/\psi N$ scattering should be dominated by the soft gluon exchange mechanism and the $S$-wave $J/\psi N$ scattering length should be $\lesssim -0.16$~fm.\footnote{We have checked that if using the single-channel Omn\`es representation and guiding the $\pi\pi$ phase shift above 0.83~GeV smoothly to $\pi$ at infinity, we would get a scattering length $\lesssim-0.3$~fm. The qualitative conclusion that the gluon exchange mechanism is dominant remains the same.}

The results obtained here can be checked using lattice QCD and lay a foundation for possible studies of the gluonic structure of the nucleon using the low-energy $J/\psi N$ scattering.\\


\noindent {\bf Note added}\\[-4mm]

After this work appeared on arXiv, we noted a lattice QCD calculation~\cite{Lyu:2024ttm} with results for $a_{J/\psi N}$ in the range of $-0.42$ and $-0.28$~fm (the sign has been converted to the convention used here), consistent the result obtained here.\\

\noindent {\bf Declaration of competing interest}\\[-4mm]

The authors declare that they have no conflicts of interest in this work.\\

\noindent {\bf Acknowledgments}\\[-4mm]

We are grateful to Xiong-Hui Cao, Hao-Jie Jing, and Mao-Jun Yan for useful discussions. This work is supported in part by the National Natural Science Foundation of China under Grants No. 12125507, No. 12361141819, and No. 12447101; and by the Chinese Academy of Sciences under Grant No.~YSBR-101.

\onecolumngrid 

\begin{appendix}

    \section{Chiral Lagrangians and amplitudes}

    The LO chiral baryon-meson Lagrangian is given by~\cite{Krause:1990xc}:
    \begin{align}
    \mathcal{L}_{\mathbb{M} \BB}^{(1)}&=\left\langle \bar{\BB}(i \slashed{\mathcal{D}} -m_0)\BB \right\rangle+\frac{D}{2}\left\langle \bar{\BB}\gamma^\mu\gamma_5 \{ u_\mu,\BB \} \right\rangle+\frac{F}{2}\left\langle\bar{\BB}\gamma^\mu\gamma_5 \left[ u_\mu,\BB \right] \right\rangle ,  \label{LO Lagrangian}
    \end{align}
    which contains three LECs, $m_0$, $D$ and $F$. 
    Here, $\langle \cdots \rangle$ means trace in the flavor space, the chiral vielbein and covariant derivative are given by: 
    \begin{align}
      u_\mu = i u^{\dagger} \partial_\mu u-i u \partial_\mu u^{\dagger},\quad \mathcal{D}_\mu \BB = \partial_\mu B+\left[\Gamma_\mu, B\right], \quad \Gamma_\mu = \frac{1}{2} \left(u^{\dagger}\partial_\mu u + u\partial_\mu u^\dagger\right),
    \end{align}
    with $u^2= U$, $U = \exp \left(i {\sqrt{2}\Phi}/{F_\pi}\right)$ with $F_\pi$ the pion decay constant, and $\BB$ and $\Phi$ are the usual ground state baryon and light pseudoscalar meson octets:
    \begin{equation}
        \mathbb{B}=\left(\begin{array}{ccc}\frac{1}{\sqrt{2}} \Sigma^0+\frac{1}{\sqrt{6}} \Lambda & \Sigma^{+} & p \\ \Sigma^{-} & -\frac{1}{\sqrt{2}} \Sigma^0+\frac{1}{\sqrt{6}} \Lambda & n \\ \Xi^{-} & \Xi^0 & -\frac{2}{\sqrt{6}} \Lambda\end{array}\right), \quad 
        \Phi=\left(\begin{array}{ccc}\frac{\pi^0}{\sqrt{2}}+\frac{\eta}{\sqrt{6}} & \pi^{+} & K^{+} \\ \pi^{-} & -\frac{\pi^0}{\sqrt{2}}+\frac{\eta}{\sqrt{6}} & K^0 \\ K^{-} & \bar{K}^0 & -\frac{2}{\sqrt{6}} \eta\end{array}\right).
    \end{equation}
    Notice that $\Gamma_\mu$ is a vector, and hence $\PP$ from this term cannot be in the $S$-wave. As a result, only the $t$- and $u$-channel exchanges depicted in Fig.~2a, b, d, e in the main text, which contribute to the left-hand cut (LHC) part of $T_{N\bar{N}\to\PP}(s)$, are present at LO for the amplitude with $S$-wave $\PP$.
    
    The $\mathcal{O}(p^2)$ chiral Lagrangian contains the $N\bar{N}\PP$ contact contribution with the $S$-wave $\PP$, as illustrated in Fig.~2c, f in the main text, and the relevant terms are~\cite{Frink:2004ic,Oller:2006yh}:
    \begin{align}
        \mathcal{L}_{\mathbb{MB}}^{(2)}=&\, b_D \langle \bar{\BB} \{ \chi_+,\BB \} \rangle+b_F \langle \bar{\BB} \left[ \chi_+,\BB \right] \rangle+b_0 \langle \bar{\BB}\BB \rangle \langle \chi_+ \rangle+b_1 \langle \bar{\BB} \left[ u^\mu ,\left[ u_\mu,\BB\right] \right] \rangle+b_2 \langle  \bar{\BB} \{ u^\mu ,\{ u_\mu ,\BB \} \} \rangle\notag\\
        &+b_3 \langle \bar{\BB} \{ u^\mu, \left[ u_\mu,\BB\right] \} \rangle+b_4\langle \bar{\BB}\BB \rangle\langle u^\mu u_\mu \rangle+ib_5 \left( \langle \bar{\BB} \left[ u^\mu,\left[ u^\nu,\gamma_\mu \mathcal{D}_\nu \BB\right]\right] \rangle - \langle \bar{\BB} \overleftarrow{\mathcal{D}}_\nu \left[ u^\nu,\left[ u^\mu,\gamma_\mu \BB\right] \right] \rangle \right) \notag \\
        &+ib_6 \left( \langle \bar{\BB} \left[ u^\mu,\{ u^\nu,\gamma_\mu \mathcal{D}_\nu \BB\} \right] \rangle - \langle \bar{\BB} \overleftarrow{\mathcal{D}}_\nu \{ u^\nu,\left[ u^\mu,\gamma_\mu \BB\right] \} \rangle \right)+ib_7 \left( \langle \bar{\BB} \{ u^\mu,\{ u^\nu,\gamma_\mu \mathcal{D}_\nu \BB\} \} \rangle - \langle \bar{\BB} \overleftarrow{\mathcal{D}}_\nu \{ u^\nu,\{ u^\mu,\gamma_\mu \BB\} \} \rangle \right)\notag\\
        &+ib_8\left( \langle \bar{\BB}\gamma_\mu \mathcal{D}_\nu \BB \rangle-\langle \bar{\BB}\overleftarrow{\mathcal{D}}_\nu\gamma_\mu \BB \rangle \right) \langle u^\mu u^\nu \rangle+\raisebox{.5ex}{\ldots} ,\label{NLO Lagrangian}
    \end{align}
    where $\chi_{ \pm}=u^{\dagger} \chi u^{\dagger} \pm u \chi^{\dagger} u, \chi=2B_0 \mathcal{M}$ with $B_0$ a constant related to the quark condensate in the chiral limit and $\mathcal{M}$ the light-quark mass matrix. In the isospin limit, the pion mass squared is $M_{\pi}^2 = 2 B_0\hat{m}$, where $\hat{m}$ represents the average mass of the $u$ and $d$ quarks.  We use the LECs values from Fit~II in Ref.~\cite{Ren:2012aj}, which are $F_\pi=87.1$~MeV, $D=0.80$, $F=0.46$, $b_D=0.222(20)$~${\rm GeV}^{-1}$, $b_F=-0.428(12)$~${\rm GeV}^{-1}$, $b_0=-0.714(21)$~${\rm GeV}^{-1}$, $b_1=0.515(132)$~${\rm GeV}^{-1}$, $b_2=0.148(48)$~${\rm GeV}^{-1}$, $b_3=-0.663(155)$ ${\rm GeV}^{-1}$, $b_4=-0.868(105)$~${\rm GeV}^{-1}$, $b_5=-0.643(246)$~${\rm GeV}^{-2}$, $b_6=-0.268(334)$~${\rm GeV}^{-2}$, $b_7=0.176(72)$~${\rm GeV}^{-2}$, $b_8=-0.0694(1638)$~${\rm GeV}^{-2}$. 
    
    For the tree-level $S$-wave amplitudes for $N\bar{N}\to\PP$, the LHC parts depicted in Fig.~2a, b and d, e in the main text read (for similar expressions in the context of calculating the effective coupling of the $\sigma$ meson to the light octet baryons, see Ref.~\cite{Wu:2023uva}):
    \begin{align}
      \hat{\mathcal A}_{N\bar{N}\to\pi\pi}(s)=&-\frac{\sqrt{3}(D+F)^2}{2F_\pi^2 }\frac{L(s,m_N,m_N,M_\pi)}{s-4m_N^2},\label{equNNpipitu}\\
      \hat{\mathcal A}_{N\bar{N}\to K\bar{K}}(s)=&-\frac{(D+3F)^2}{12F_\pi^2}\frac{L(s,m_N,m_\Lambda,M_K)}{s-4m_N^2}-\frac{3(D-F)^2}{4F_\pi^2}\frac{L(s,m_N,m_\Sigma,M_K)}{s-4m_N^2} ,\label{equNNKKbartu}
    \end{align}
    where $m_N$, $M_\pi$, $m_\Lambda$, $M_K$, and $m_\Sigma$ are the isospin-averaged masses of the corresponding particles, $s$ is the $N\bar N$ invariant mass squared, and
    \begin{align} 
    L(s,m_1,m_2,m)&=(m_1+m_2)\left[s-2m_1(m_1-m_2)+H_0(s,m_1,m_2,m)H_1(s,m_1,m_2,m)\right] \,,\notag\\
    H_0(s,m_1,m_2,m)&=2(m_1+m_2)\left[-2m^2m_1+2m_1(m_1-m_2)^2+m_2s \right] ,\notag\\
    H_1(s,m_1,m_2,m)&=\frac{H_2^+(s,m_1,m_2,m)-H_2^-(s,m_1,m_2,m)}{2\sqrt{(s-4m^2)(s-4m_1^2)}}\,,\notag\\
    H_2^\pm(s,m_1,m_2,m)&={\rm{ln}}\left[ s-2(m^2+m_1^2-m_2^2) \mp \sqrt{(s-4m^2)(s-4m_1^2)} \right] .\notag
    \end{align}
    The contact terms, which are from the NLO Lagrangian, read:
    \begin{align}
        \mathcal A_{N\bar{N}\to\pi\pi}(s)&=\frac{1}{4 \sqrt{3} F_\pi^2}\Bigl(8 M_\pi^2(6 b_0-3(b_1+b_2+b_3+2 b_4-b_D-b_F)-2(b_5+b_6+b_7+2 b_8) m_N)\notag\\
        &\quad+4(3(b_1+b_2+b_3+2 b_4)+(b_5+b_6+b_7+2 b_8) m_N) s\Bigr),\label{equNNpipis}\\
        \mathcal A_{N\bar{N}\to K\bar{K}}(s)&=\frac{1}{3 F_\pi^2}\Bigl( 2M_K^2(12b_0-3(3b_1+3b_2-b_3+4b_4-3b_D+b_F)+2(-3b_5+b_6-3b_7-4b_8)m_N)\notag\\
        &\quad +(9b_1+9b_2-3b_3+12b_4+(3b_5-b_6+3b_7+4b_8)m_N)s\Bigr)  .\label{equNNKKbars}
    \end{align} 
    
    It is important to emphasize that the aforementioned tree-level $S$-wave amplitudes, specifically Eqs.~(\ref{equNNpipitu}-\ref{equNNKKbars}), are free of kinematic singularities~\cite{Martin:1970hmp}.  This is achieved by dividing the amplitudes by a factor $\sqrt{s-4m_N^2}$~\cite{Wu:2023uva}, thereby rendering them suitable for subsequent analysis using the Muskhelishvili-Omn\`es representation. Furthermore, we implement partial-wave expansions for the helicity amplitudes (for a detailed introduction, see Ref.~\cite{Martin:1970hmp}), and the components associated with polarization satisfy:
    \begin{align}
        \hat{\mathcal A}_{N(\lambda_3)\bar{N}(\lambda_4)\to\PP}(s)=(\lambda_3+\lambda_4)\hat{\mathcal A}_{N\bar{N}\to\PP}(s), \quad
        \mathcal A_{N(\lambda_3)\bar{N}(\lambda_4)\to\PP}(s)=(\lambda_3+\lambda_4)\mathcal A_{N\bar{N}\to\PP}(s),
    \end{align}
    where $\lambda_3$ and $\lambda_4$ denote the third components of helicities for the nucleon and antinucleon, respectively. Thus, Eqs.~(\ref{equNNpipitu}-\ref{equNNKKbars}) can be viewed as the tree-level $S$-wave amplitudes for $\lambda_3=\lambda_4={1}/{2}$.

    The chiral Lagrangian for the $\psi_i\to\psi_j\PP$ transition reads~\cite{Mannel:1995jt,Chen:2015jgl}:
    \begin{align}
    \mathcal{L}_{\psi}=  \vec \psi_j^\dag \cdot \vec \psi_i \left(c_1^{(ij)} \left\langle {u_{\mu} u^{\mu}}\right\rangle+c_2^{(ij)}\left\langle {u_{\mu} u_{\nu}}\right\rangle v^{\mu} v^{\nu}+c_{m}^{(ij)}  \left\langle \chi_+\right\rangle\right)+{\rm H.c.}\label{cheft},
    \end{align}
    where $\vec{\psi}_1$ ($\vec{\psi}_2$) annihilates the $J/\psi$ ($\psi(2S)$), $v_\mu$ is the four-velocity of the charmonia, and $c_{1,2,m}^{(ij)}$ are LECs.
    
     The $\psi_i\to\psi_j\PP$  transition amplitude, decomposed into a sum of $S$ and $D$ waves, reads:
    \begin{align}
    {\cal A}^{\rm tree}_\mathcal{P}(s,\theta_1)=&\,{\cal A}^S(s,M_{\mathcal P})+{\cal A}^D(s,M_{\mathcal P})P_2(\cos\theta_{\mathcal P})\,,\\
     {\cal A}^S(s,M_{\mathcal P})=&\,-\frac{2I_{\mathcal{P}}}{F_\pi^2}\left\{c_1^{(ij)}\left(s-2 M_{\mathcal P}^2\right)+2c_m^{(ij)}M_{\mathcal P}^2+\frac{c_2^{(ij)}}{2}\left[s+\bm{q}^2\left(1-\frac{\sigma_{\mathcal P}^2}{3}\right)\right]\right\}, \\
     {\cal A}^D(s,M_{\mathcal P})=&\,\frac{2I_{\mathcal{P}}}{3 F_\pi^2} c_2^{(ij)} \bm{q}^2 \sigma_{\mathcal P}^2,
    \end{align}
    where $\theta_{\mathcal P}$ is the angle between the 3-momentum of the ${\mathcal P}$ in the rest frame of the $\PP$ system and that of the $\PP$ system in the rest frame of the initial $\psi_{i}$, $P_2(\cos\theta_{\mathcal P})$ denotes the second-order Legendre polynomial, $I_{\pi}=\sqrt{3/2}$ and $I_{K}=\sqrt{{2}}$ represent the isospin factors, $\bm{q}$ is the 3-momentum of the final charmonium in the rest frame of the initial state, and $\sigma_{\mathcal P} \equiv \sqrt{1-4 M_{\mathcal P}^2/s}$.
    
    In Ref.~\cite{Dong:2021lkh}, $\alpha^{(21)}$ was obtained by fitting the BESII experimental data on the $\psi(2S)\to J/\psi\pi^+\pi^-$ transition~\cite{BES:2006eer}. However, the operator $\left\langle \chi_+\right\rangle$ was not included there.
    We amend the flaw and update the fit, with additional ATLAS data~\cite{ATLAS:2016kwu}. The fitting results are shown in Fig.~3 in the main text.
    
    We use two sets of coupled-channel Omn\`es matrix from Refs.~\cite{Ropertz:2018stk,Hoferichter:2012wf}, and find similar fitting results with $\chi^2/{\rm d.o.f.}\simeq1.5$. 
    The fitted parameters read:
    \begin{align}
    |\alpha^{(21)}|=(1.18\pm0.01\pm0.05)\,{\rm GeV^{-3}},\ \kappa=0.26\pm0.01\pm0.01,\label{eq:betakappa}
    \end{align}
    where the first uncertainties come from the errors of the experimental data, while the second ones represent the difference from the two values of the Omn\`es matrix. 
        
\end{appendix}

\newpage

\section*{References}
\twocolumngrid 

\bibliographystyle{fundamentalresearch} 
\bibliography{refs}

\end{document}